\documentclass[aps,amsmath,twocolumn,amssymb,floatfix,showpacs,
superscriptaddress,footinbib,longbibliography,prb,reprint]{revtex4-1}
\pdfoutput=1
\usepackage{graphics}
\usepackage{bm}
\usepackage{epsfig}
\usepackage{enumerate}
\usepackage{subfigure}
\usepackage{amsmath}
\usepackage{color}
\usepackage{braket}
\usepackage{graphicx}
\usepackage{hyperref}
\hypersetup{
    colorlinks=true,
    linktoc=page,
    linkcolor=red,
    citecolor=blue,
    urlcolor=magenta
}
\usepackage[utf8]{inputenc}

\newcommand{\trt}   {TaRhTe$_4$}
\newcommand{\tit}   {TaIrTe$_4$}
\newcommand{\nit}   {NbIrTe$_4$}

\begin{document}

\title{The orbital-driven topological phase transition and  planar Hall responses in ternary tellurides Weyl semi-metals}

\author{Banasree Sadhukhan}
\email[Corresponding author:]{banasree.sadhukhan@mahindrauniversity.edu.in}
\affiliation{Department of Physics, École Centrale School of Engineering,  Mahindra University,  Hyderabad,  Telangana 500043,  India}
\author{Tanay Nag}
\email{tanay.nag@hyderabad.bits-pilani.ac.in}
\affiliation{Department of Physics, BITS Pilani-Hydrabad Campus, Telangana 500078, India}

\date{\today}

\begin{abstract}
We study electronic properties of the ternary tellurides TaXTe$_4$ (X=Rh, Ir) using density functional theory and investigate chiral anomaly mediated planar Hall response from ab initio calculations. We show that  TaRhTe$_4$ is a hybrid Weyl semimetal (WSM), hosting Weyl points (WPs) of both type-I,  type-II, and TaIrTe$_4$ is a type-I WSM in absence of spin-orbit coupling (SOC). TaRhTe$_4$ continues to remain a hybrid WSM  while 
TaIrTe$_4$ converts into a type-II WSM  under the application of SOC.   We observe long Fermi arcs connecting WPs of opposite chirality.  We report orbital-driven topological phase transition in ternary tellurides. {{The WSM phases in TaXTe$_4$ are controlled by the orbital character of the $d_{xz}$ and $d_{z^2}$ states of X=Ir/Rh atoms.  Replacing Rh with Ir enhances the $d_{z^2}$ orbital contribution near the Fermi level at the expense of $d_{xz}$ states. This transforms the type-I WPs into type-II resulting in a conversion of  hybrid WSM TaRhTe$_4$ to type-II WSM TaIrTe$_4$. }}   This systematic study opens new routes for engineering topological materials relying beyond strong SOC and sheds light on {{ the effect of orbital degree of freedom}} on the electronic properties of tellurides.  We further report an enhancement of planar Hall effects due to orbital-driven topological phase transition in TaXTe$_4$ and  we make resort to a tight-binding model to correlate the above findings with the {{velocity modulated off-diagonal effective mass anisotropy}} in different types of WSMs.

 \end{abstract}


\maketitle


\section{Introduction} \label{intro}

 Topological systems range from insulating phases \cite{RevModPhys.82.3045}, where the valence and conduction bands are separated by an energy gap, to semimetallic phases \cite{RevModPhys.90.015001}, where these bands intersect at discrete points or lines in momentum space. The former one is referred to as topological insulator while the later is dubbed as Dirac and  Weyl semimetals (WSMs). The defining feature of the WSMs is the crossing of linearly dispersing non-degenerate (bulk) bands at a single point in the momentum $k$-space, known as Weyl points (WPs), which requires breaking of either time reversal or inversion symmetry.  The elementary quasi particle excitations around the WPs are, therefore, massless chiral fermions (Weyl fermions) and obey Weyl equation \cite{balents2011,vishwanath2015}. The topology of the electronic band structure ensures that, although WPs can be created or annihilated by symmetry preserving perturbations \cite{ghimire2019}, the WSM phase cannot be gapped out.  Moreover, the WSMs host an exotic surface state band structure containing topological Fermi arcs terminating at the projection of bulk  WPs of opposite chirality.  These topological states of matter are being investigated for exotic transport properties \cite{PhysRevB.84.075129,de2017quantized,PhysRevB.104.245122,PhysRevLett.107.127205,PhysRevB.93.035116,zhang2020spin,wu2017giant,PhysRevLett.117.077202,lu2017quantum,PhysRevB.107.L081110,PhysRevB.106.045424,RevModPhys.90.015001,CULCER2012860,GUSEV2019113701,tkachov2013spin,PhysRevB.107.245141,PhysRevB.104.115420,PhysRevLett.123.216802,lai2021third,PhysRevB.103.144308, PhysRevB.100.155144,Barreto2014}.  A notable transport property of quantum materials such as Dirac or WSMs is negative longitudinal magneto resistivity,  which is often labelled as a signature of chiral anomaly \cite{PhysRevB.88.104412,  PhysRevLett.113.247203,  PhysRevLett.117.136602,  Hirschberger2016,   PhysRevX.5.031023,  NIELSEN1983389,  doi:10.1126/science.aac6089}.

Two types of WSMs with distinct band structure have been reported and experimentally discovered in real materials.  The conventional type-I WSMs are characterized by shrinking of the Fermi surface to a point at the WP energy. The simplest Fermi surface of such a WSM would consist of only two such points. The family of materials MX (M = Ta and Nb, and X = As and P) were theoretically predicted and later experimentally confirmed to be type-I WSMs \cite{lv2015a,xu2015a,yang2015,xu2015b,weng2015,lv2015b,xu2015c,sun2015}.  On the other hand, in type-II WSMs, which are realized only in condensed matter systems and have no equivalent in high energy physics, the WPs are tilted and appear as a connector of electron and hole pockets \cite{soluyanov2015}.  This large tilt of the cone leads to finite density of states at the WP energy. As a result, type-II WSMs exhibit different physical properties as compared to the type-I WSMs.  There exist a new type quantum materials namely,  hybrid WSM which combines both type-I and type-II WPs but detailed quantitative studies are lacking \cite{PhysRevB.94.121105,  ALISULTANOV20183211,  Xu_2024}.  Therefore,  finding new candidate for hybrid WSMs and study its topological transport have drawn tremendous attention recently for its potential applications \cite{PhysRevLett.121.136401,   PhysRevB.107.195426,  Li2023}.

Now coming to the semi classical transport of WSMs,  chiral anomaly is caused by the pumping of charge carriers between two  WPs with opposite chirality when both electric and magnetic fields are parallel.  Chiral anomaly and non-trivial Berry curvature  together induce another key effect,  the Planar Hall effect (PHE) \cite{PhysRevLett.119.176804,  PhysRevResearch.2.022029,  PhysRevB.100.205128,  PhysRevB.99.115121,  PhysRevB.98.161110,  PhysRevB.98.121108,  PhysRevB.98.041103,  PhysRevB.96.041110}.  In a PHE arrangement,  the electric and magnetic fields are coplanar rather than mutually perpendicular,  as in a conventional Hall effect.  In addition to the chiral anomaly origin of PHE,  it is also observed in ferromagnetic materials originating from strong spin-orbit coupling (SOC) \cite{PhysRevLett.90.107201,  PhysRevB.78.212402}.  In non-magnetic semimetals, this effect can also be induced by the orbital anisotropy of the electronic structure and Fermi surface \cite{PhysRevResearch.2.022029,  PhysRevX.8.031002}.  In topological semimetals, the anisotropic orbital magneto resistance can also contribute significantly to PHE \cite{PhysRevMaterials.3.014201,  PhysRevB.99.155119,   liang2019origin,   Meng_2020}.  It is noteworthy that PHE behaves quite differently in type-II WSMs as compared with type-I semimetals as the band crossings of electron pockets and hole pockets are tilted in type-II WSMs. Therefore,  it has become important to investigate PHE in different topological phases,  as evidenced in {{ a recent study \cite{PhysRevB.107.L081110} }}. Importantly,  WSMs in the form of binary and ternary tellurides have attracted a lot of attention recently as they exhibit topological Fermi arcs \cite{li2017,wu2016, tamai2016,deng2016,haubold2017,belopolski2017}, exotic transport phenomena such as large magneto resistance \cite{ali2014,jiang2015,khim2016,thirupathaiah2017}, and superconductivity \cite{cai2019,lu2016,qi2016,xing2018}.

The different topological states and chirality mediated transport can be understood by the role of the {{off-diagonal effective mass \cite{PhysRevB.105.205207,  PhysRevB.108.235136}}} of the system of interest. The Berry curvature contributes to anomalous velocity which is modified by band curvature in addition to the band dispersion.  For example,  WSMs exhibit chiral anomaly-driven transport that depends on the band curvature which is intrinsically related to the effective mass of the quasi particles near the WPs.   Structural modification leads to mass anisotropy which is found to be  responsible for realizing hybrid nodal-line fermion state in transition-metal-tetraphosphide \cite{PhysRevB.108.235136}. Interestingly,   how this effective mass affects the chirality mediated transport in topological materials has not been investigated yet \cite{PhysRevB.105.205207}.

In the current study,   we investigate the orbital-driven topological phase transition from type-II to hybrid WSM and its effect on the planar Hall transport in ternary tellurides.  We report the effect of SOC in ternary tellurides and the topological phase transitions there which are essentially driven by orbital contribution.  Without SOC,  TaIrTe$_4$ appears as type-I WSM which transforms into type-II phase with SOC.  If we replace Ir by an element from the same isoelectronic group, namely Rh, then 
TaRhTe$_4$ emerges as a hybrid WSM which host type-I and type-II WPs simultaneously. Note that TaRhTe$_4$ is  always in hybrid WSM phase independent of SOC.

WSM phase in both ternary tellurides originates from d orbitals of Ta atoms Ta-${5d_{z^2}}$,  -${5d_{xz}}$,  and p orbitals of Te atoms Te-${5p_{y}}$ where the nature of d orbital symmetry play an crucial role in determining type of WPs.  {{In TaRhTe$_4$,  the WPs below the Fermi level are of type-II in nature,  whereas WPs above Fermi level are of type-I in nature.  Both type of WPs in TaRhTe$_4$ are contributed from Rh-${4d_{xz}}$ and -4${d_{z^2}}$ which makes TaRhTe$_4$ as hybrid WSM.  The substitution of Rh by Ir induces a redistribution of orbital occupancy, leading to an enhanced contribution of the Ir-5${d_{z^2}}$ orbital relative to the Ir-5${d_{xz}}$  near the Fermi level.  This orbital reorganization increases the tilting of the type-I WPs and eventually converts them to type-II WPs above Fermi level while already existing type-II WPs below Fermi level are annihilated. As a result, TaIrTe$_4$ emerges as type-II WSM  without altering the underlying crystal symmetry as compared to TaRhTe$_4$.}} Additionally,  we also systematically study the effect of topological phase on PHE in both TaRhTe$_4$ and TaIrTe$_4$.  Finally we employed a tight-binding model for type-I,  type-II and hybrid WSM to explain our findings in connection to its  {{velocity modulated off-diagonal effective mass.}}

 The manuscript is organised as follows.  In Sec.~\ref{sec1},  we investigate topological characterization of WPs in TaXTe$_4$ (X=Rh, Ir) followed by  Sec.~\ref{sec2} where  we discuss our results on the electronic structure and topological properties.  In Sec.~\ref{sec3},  next we explore how the orbital-driven topological phase transition affects chiral anomaly mediated planar Hall responses of both ternary tellurides and employ a tight-binding model to investigate the role of the {{velocity modulated off-diagonal effective mass anisotropy}} on PHE in Sec.~\ref{sec4}.  Finally, in Sec.~\ref{sec5},  we summarize our results and end with a conclusion.


\begin{figure}[ht]
\centering
\includegraphics[width=0.48\textwidth,angle=0]{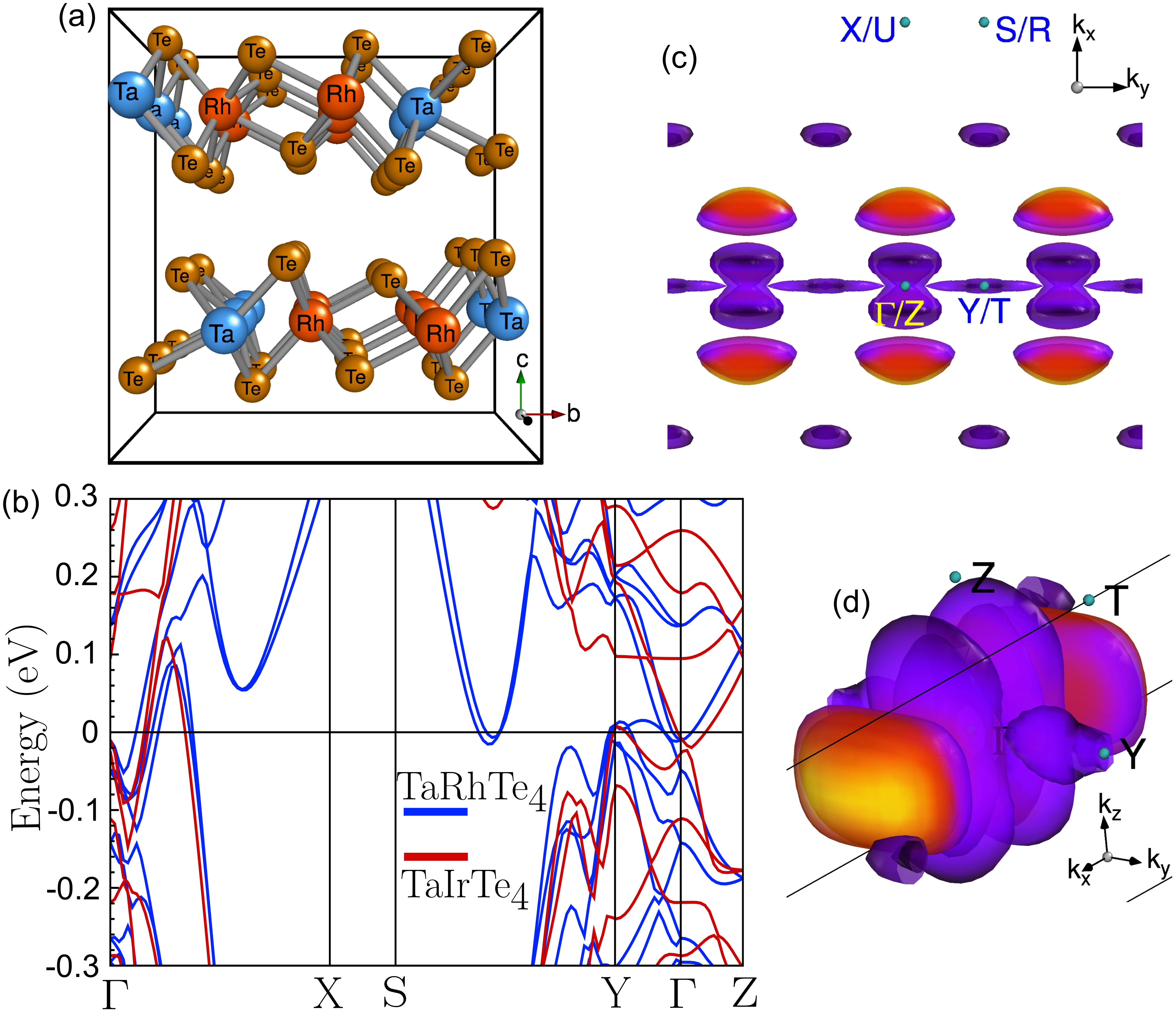}
\caption{(a) Crystal structure of TaRhTe$_4$ (TaIrTe$_4$) showing the layered
        structure, where Ta, Rh (Ir) and Te atoms are shown by 
        by blue, orange and brown filled spheres, respectively. (b) The
        full relativistic band structures of  TaRhTe$_4$ and TaIrTe$_4$ are depicted in blue and red color, respectively.
        (c) $k_y$-$k_x$ projected, and (d) three dimensional view of the bulk
        Fermi surface of TaRhTe$_4$.}
\label{fig1}
\end{figure}


\section{Effect of SOC and topological characterization of Weyl points in ternary tellurides}
\label{sec1}

 The orthorhombic $T_d$-TaXTe$_4$ (X=Rh,  Ir)  crystallizes in a layered structure with space group $Pmn2_1$ (No. 31) and is shown in Fig.  \ref{fig1}(a). The layered structure can be described by a unit cell containing four formula units which form two layers (along the $c$-axis). The corresponding lattice parameters are: $a = 3.75672$ (3.77) {\AA}, $b = 12.5476$ (12.421) {\AA}, and $c = 13.166$ (13.184) {\AA} for TaRhTe$_4$ (TaIrTe$_4$).   Within each layer, the transition metal ions are octahedrally coordinated by Te,  leading to a network of distorted edge-sharing MTe$_6$ (M = transition metal) octahedron. The crystal structure lacks inversion symmetry $\mathcal{P}$, but posses a mirror symmetry $m(x)$,  a glide mirror symmetry $\{m(y)|\bf{t}\}$ and a glide rotational twofold symmetry about the $z$ axis $\{c_2(z)|\bf{t}\}$ where ${\bf{t}}(\frac{1}{2}, 0, \frac{1}{2})$ is a fractional translation of a Bravais lattice vector.These materials preserve time reversal symmetry due to their non-magnetic structures.

The electronic band structure along $\Gamma$-$X$ without SOC and with SOC cases are found to be similar for both isostructural and isoelectronic compounds TaRhTe$_4$ and TaIrTe$_4$, see Fig. \ref{fig1}(b).  However, there are significant difference, such as the presence of an additional electron pocket along the path $S$-$Y$, and absence of a hole pocket at the $Y$ point for TaRhTe$_4$  \cite{SM}.  Upon inclusion of SOC, each band splits into two almost everywhere in BZ. Figure \ref{fig1}(b) shows the band structure in a small energy window around the Fermi energy for TaXTe$_4$ (X=Rh, Ir).  There are four partially filled bands that give rise to a variety of electron and hole pockets.  Consequently, the bulk Fermi surface [see Figs.  \ref{fig1}(c) and \ref{fig1}(d)] consists of nested electron and hole pockets along $\Gamma$-$X$, small electron pockets along $Y$-$S$, and small hole pockets along ($Y$-$\Gamma$) for TaRhTe$_4$.   Note that all pockets are nested due to small spin-orbit splitting of the  bands.  In particular, the conduction bands around the $\Gamma$ point lead to three-dimensional (3D) electron pockets, while the valence bands lead to two nested but not touching hole pockets along $\Gamma$-$X$.  The small electron pockets along $Y$-$S$ also consist of nested surfaces.  The hole pocket along $\Gamma$-$Y$ has an open Fermi surface across the $k_y = \pm \pi$ BZ boundary, leading to a quasi-two-dimensional dispersion.  These features of the bulk Fermi surface are significantly distinct from {\tit} despite the structural similarity due to different band dispersion \cite{klaus2016} as shown in Fig. \ref{fig1}(b).

\begin{figure}[ht]
\centering
\includegraphics[width=0.5\textwidth,angle=0]{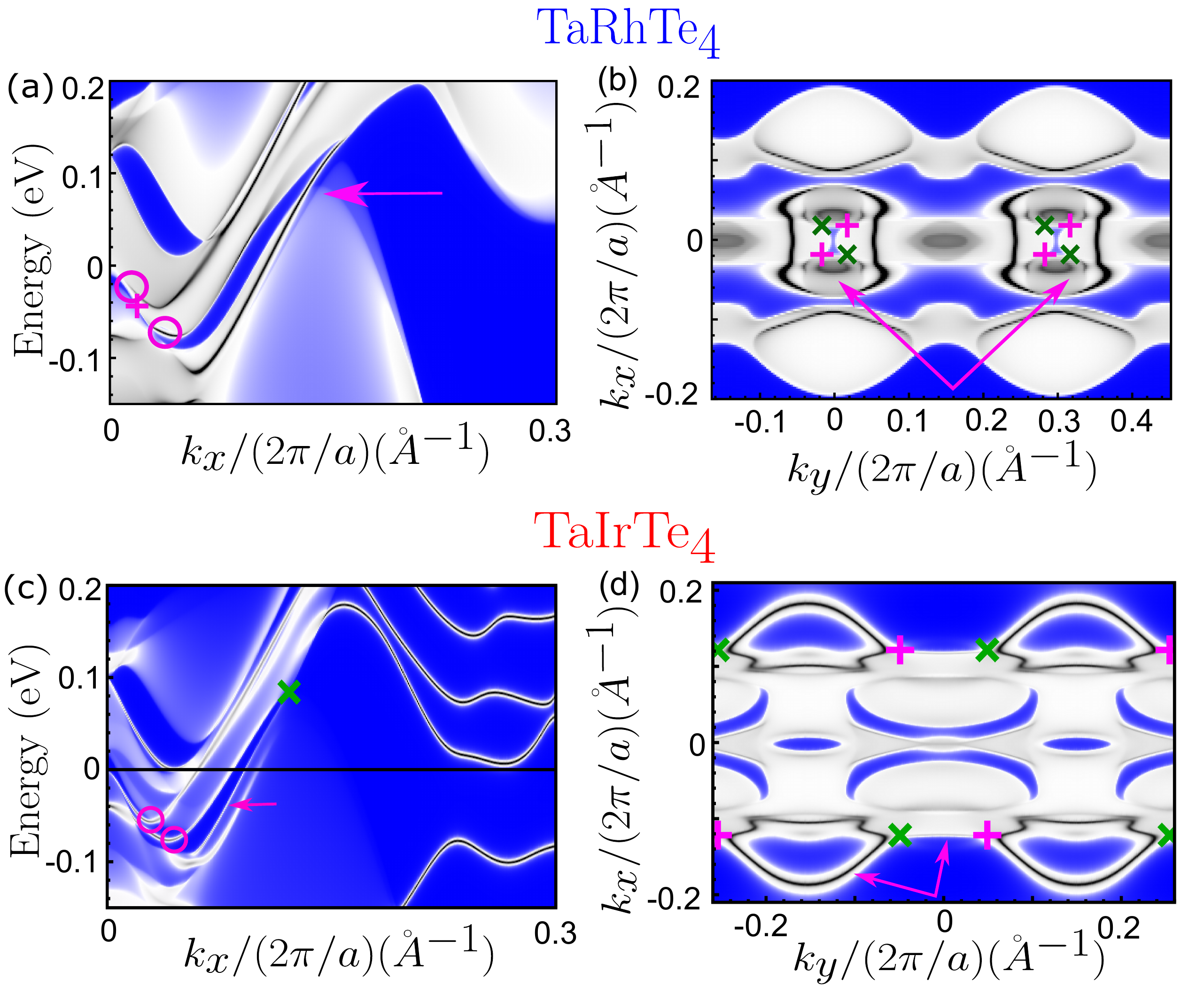}
\caption{Spectral function for  
    (001) surface in (a) and for (00$\bar{1}$) surface in (b)  along the line $\frac{2\pi}{a}$($k_x$,$k_y^{\rm
    WP}$), capturing W$_1$ respectively in TaRhTe$_4$.  Fermi surface maps for (001) surface in (c) and  (00$\bar{1}$) surface in (d) at $E$ = $E^{W_1}$ respectively in TaRhTe$_4$.   Fermi arcs i.e., topological surface states are marked by arrows while the trivial surface states are
    marked by circles. The position of the WP is shown with the symbols '+' 
    and '$\times$', respectively, for chirality +1 and -1. 
}
    \label{fig3}
\end{figure}

The topological properties without SOC are found to be quite similar between the two ternary tellurides \cite{SM}.  TaRhTe$_4$ (TaIrTe$_4$) appears as hybrid (type-I) WSM without SOC.   WPs in TaRhTe$_4$ are labelled as V$_1$ and V$_2$, whereas they are labelled as U$_1$  and U$_2$ for TaIrTe$_4$, as demonstrated in table \ref{table2} and table  \ref{table1} for TaIrTe$_4$, and TaRhTe$_4$, respectively. 
V$_1$ appears as type-II in nature and V$_2$ appears as type-I in nature which makes TaRhTe$_4$ as hybrid WSM without SOC. On the other hand, TaIrTe$_4$ is a type-I WSM without SOC because of type-I nature of both the WPs  U$_1$  and U$_2$.
All WPs in TaXTe$_4$ are fourfold degenerate obeying symmetry of the system ($E$,  $m(x)$,  $\{m(y)|\bf{t}\}$ and $\{c_2(z)|\bf{t}\}$ where $E$ is identity term) and are presented in supplementary materials \cite{SM,  klaus99,  PW92,  fplo_web,  klaus2016,  PhysRevB.103.144308}.  In the presence of SOC, TaRhTe$_4$ remains a  hybrid WSM even after new type-II WPs are created. By contrast,  TaIrTe$_4$ converts into a type-II WSM where all type-I WPs are annihilated.

\begin{table}[t!]
    \small
{{ \caption{Types,  positions,  and energies of the WPs without and with SOC in TaRhTe$_4$. It remains a hybrid WSM irrespective of the SOC.}
    \begin{tabular*}{0.5\textwidth}{ p{1.8cm} p{4.0 cm} p {1.0 cm} p{1.3 cm} }
    \hline\hline
        WP (type)  & Position [($k_x$,$k_y$,$k_z$)$/\frac{2\pi}{a}$ ]  &  SOC  & $E$ (meV)  \\
    \hline
     V$_1$ (II) & ($\pm 0.0269, \pm 0.0054, 0$) & w/o  & $-117.6$  \\  
     V$_2$  (I)  & ($\pm 0.1505, \pm 0.0089, 0$) & w/o &  $\,\,\,\,194.0$ \\ 
     \hline  
         W$_1$  (II)  & ($\pm 0.0188, \pm 0.0172, 0$) & w &  $-43.7$  \\ \hline
          W$_2$  (II)  & ($\pm 0.0266, \pm 0.0219, 0$) & w &  $-60.8$ \\ \hline
          W$_3$  (I)  & ($\pm 0.1541, \pm 0.0354, 0$) & w &  $\,\,\,\,117.2$ \\ 
          \hline
    \hline
    \end{tabular*}
    \label{table2}}}
\end{table}

\begin{table}[ht]
    \small
     {{ \caption{Types,  positions and energies of the WPs for SOC driven topological phase transition in TaIrTe$_4$.  Without SOC,  TaIrTe$_4$ is a type-I WSM which became type-II WSM with inclusion of SOC.}
    \begin{tabular*}{0.5\textwidth}{ p{1.8cm} p{4.0 cm} p {1.0 cm} p{1.3 cm} }
    \hline\hline
        WP (type)  & Position [($k_x$,$k_y$,$k_z$)$/\frac{2\pi}{a}$ ]  &  SOC  & $E$ (meV)  \\
    \hline
      U$_1$ (I) & $(\pm 0.1178, \pm  0.0467,  0)$ & w/o & $ 116.3 $    \\
      U$_2$ (I) & $(\pm 0.0483,  \pm 0.0699,  0)$ & w/o & $ -78 $   \\ \hline
     W (II) & ($\pm 0.1217, \pm 0.0491, 0$) & w & $\,\,\,\,84.16$  \\
    \hline
    \hline
    \end{tabular*}
    \label{table1}}}
\end{table}

{{TaRhTe$_4$ hosts eight and twelve WPs in the absence and presence of SOC, respectively. The  WPs V$_1$ and V$_2$, lying below and above the Fermi level, respectively, persist under SOC and are displaced to new energy and momentum locations,  labelled as W$_2$ and W$_3$,  respectively. Note that the WPs below [above] Fermi level remain type-II [type-I] irrespective of SOC in TaRhTe$_4$.
In addition, an extra set of type-II WPs, denoted by W$_1$,  are generated below the Fermi level due to the inclusion of SOC in TaRhTe$_4$ (see Fig.\ref{fig2}(a)).}}
These WPs in TaRhTe$_4$ lie in the $k_z=0$ plane while remaining well-separated in energy and momentum space, indicating that the WSM phase in {\trt} is robust.  The coexistence of both type-I and type-II WPs makes TaRhTe$_4$ as hybrid WSM.    {{In TaIrTe$_4$,  a total of eight WPs (U$_1$ and U$_2$) are present in the absence of SOC due to symmetry of the system.  Upon inclusion of SOC, the type-I WPs U$_1$, lying above Fermi level, survive but are shifted in both energy and momentum space, appearing as the type-II WPs W (see Fig.\ref{fig2}(a)),  while the another pair of type-I WPs U$_2$, appearing below the Fermi level, annihilate during this process. }}

The details of computational part,  WPs characterization and spectral properties for both ternary tellurides are described in supplementary materials \cite{SM}.  It is interesting to note that coexistence of both type-I and type-II WPs has also been predicted in {\nit} \cite{li2017_nit,zhou2019}.   The spectral densities and Fermi surface mapping for one energy set of WPs in both TaRhTe$_4$ and TaIrTe$_4$ are presented in Fig.\ref{fig3} where long Fermi arcs are clearly visible connecting WPs with opposite chirality.



\section{Orbital driven topological phase transition}
\label{sec2}

\begin{table}[ht]
{ {\small
   \caption{Types,  positions and energies of the WPs for orbital driven topological phase transition in TaXTe$_4$ (X=Rh, Ir).  TaIrTe$_4$ is a type II WSM whereas TaRhTe$_4$ is a hybrid WSM due to coexistence of both type-I and II WPs.}
   \begin{tabular*}{0.5\textwidth}{ p{2.0cm} p{4.0 cm} p {1.2 cm} p{1.2 cm} }
    \hline\hline
        WP (type)  & Position [($k_x$,$k_y$,$k_z$)$/\frac{2\pi}{a}$ ]  & X & $E$ (meV)  \\
    \hline
    W (II) & ($0.1217, 0.0491, 0$) & Ir & $\,\,\,\,84.16$ \\ \hline
     W$_1$  (II)  & ($0.0188, 0.0172, 0$) & Rh & $-43.7$  \\
     W$_2$  (II)  & ($0.0266, 0.0219, 0$) & Rh & $-60.8$ \\
     W$_3$  (I)  & ($0.1541, 0.0354, 0$) & Rh & $\,\,\,\,117.2$ \\ 
         \hline
    \hline
    \end{tabular*}
    \label{table3}}}
\end{table}

\begin{figure}[ht]
{ {\centering
\includegraphics[width=0.5\textwidth,angle=0]{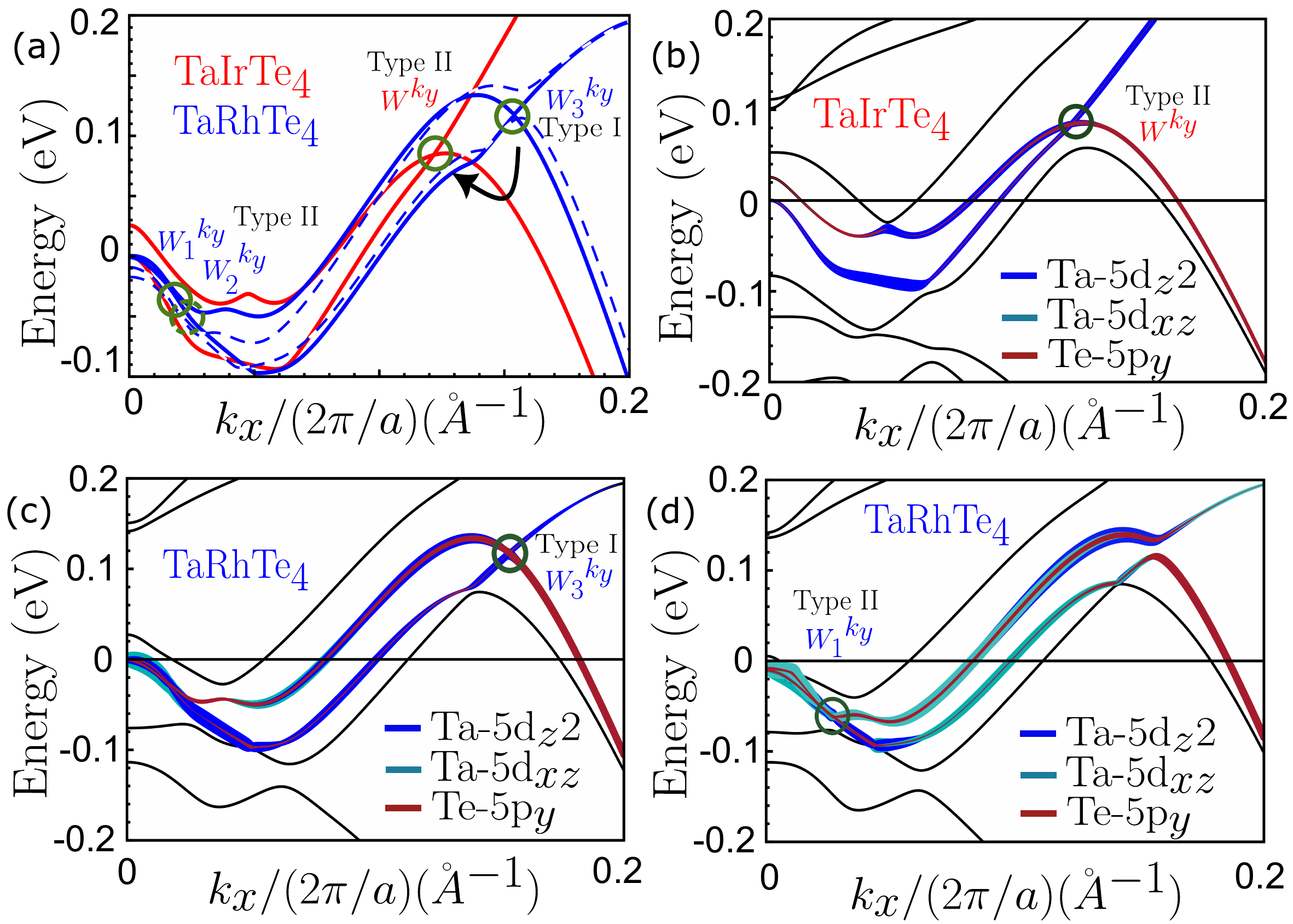}
\caption{(a) Topological phase transition from hybrid to type-II WSMs in TaXTe$_4$ (X=Ir and Rh).  The blue and red colors correspond to TaRhTe$_4$ and TaIrTe$_4$, respectively.  WPs  W$_1$ (solid line) and  W$_2$ (dotted line) in TaRhTe$_4$ are of type-II,  whereas W$_3$ (solid line) is of type-I nature which makes TaRhTe$_4$ a hybrid WSM.  W (solid line) in TaIrTe$_4$ is of type-II nature.  The type-I nature of the WP W$_3$ in TaRhTe$_4$ transforms into type-II nature of the WP W when Rh is replaced with Ir as seen in TaIrTe$_4$.  Orbital contributions of Ta and Te atoms to the WPs in (b)TaIrTe$_4$ and (c)-(d) TaRhTe$_4$ respectively. 
     }
    \label{fig2}}}
\end{figure}

We now discuss the origin of the above phase transition from viewpoint of orbital contribution. At the outset, we  note that orbital projected band inversion leads to Chern insulator resulting in the  Hall response that is driven by orbital angular momentum \cite{yao2025topological,  Gao2023}.  It opens new insights into the role of the orbital degree of freedom in topological phases of quantum matter for the exploration of orbital-driven topological phenomena beyond relying on strong SOC.  The orbital-driven topological phase transitions are generically induced by the change in orbital occupancy,  character,  symmetry,  or order that can drastically modify the electronic band structure \cite{zhang2019topological,  PhysRevB.106.125112,  guo2023correlation,  Herbrych2021}.  It is relevant for materials with strong correlations where orbital physics dominates \cite{PhysRevA.100.043608}.

\begin{figure}[ht]
\centering
\includegraphics[width=0.5\textwidth,angle=0]{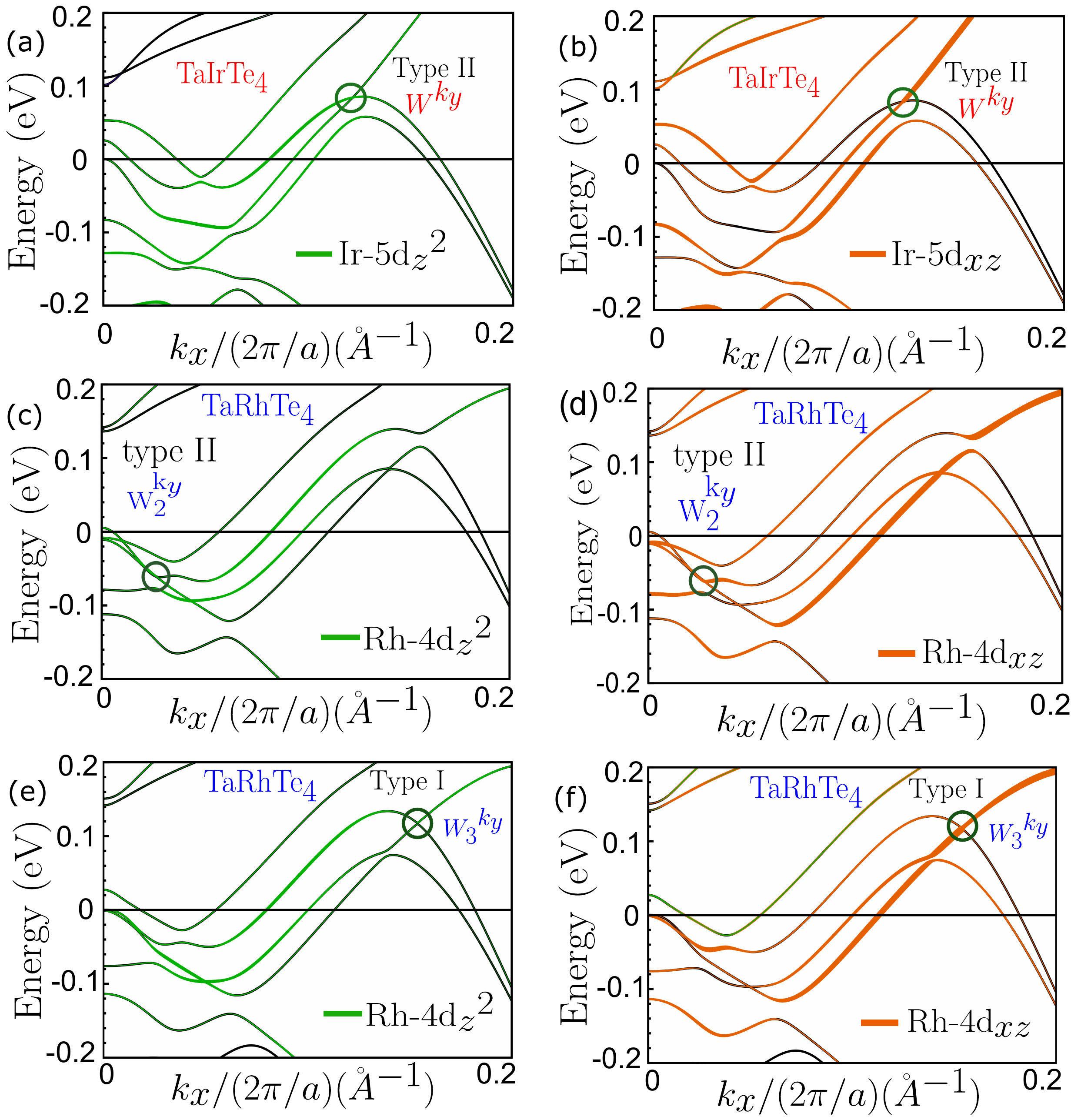}
\caption{{{ Contribution of ${d_{z^2}}$ and ${d_{xz}}$ orbitals on band crossings around WPs in (a)-(b) TaIrTe$_4$ for W (type-II) and, (c)-(f) TaRhTe$_4$ for W$_2$ (type-II)  and W$_3$ (type-I)  respectively.  Similar behaviour is also observed for another  band crossing around WP W$_1$ (type-II) in TaRhTe$_4$.}}}
    \label{fignew1}
\end{figure}

{ {In the presence of SOC, we here explore the orbital-driven topological phase transition from type-II to hybrid WSM in ternary tellurides as presented in table \ref{table3}.  The f orbitals of Ir ([Xe]4f$^{14}$5d$^{7}$6s$^{2}$) have no contributions to the  band crossings in the vicinity of WP for TaIrTe$_4$ as they form core orbitals.   In both TaIrTe$_4$ and TaRhTe$_4$,  WPs are dominantly contributed from d orbitals of Ta atoms namely Ta-${5d_{z^2}}$,  -${5d_{xz}}$,  which strongly hybridize with p orbitals of Te atoms namely Te-${5p_{y}}$ as shown in Fig.\ref{fig2}(b)-(d).  To explore orbital driven topological phase transition in TaXTe$_4$, we investigate the contribution of different d-orbitals of Rh/Ir atoms on WPs in TaXTe$_4$ (X=Rh, Ir).  }}

{ {Our orbital-resolved analysis reveals that both the { {Rh-4$d_{xz}$ and -4$d_{z^2}$}} orbitals play dominant roles in forming the  band crossings at WPs in TaRhTe$_4$.  In contrast, for TaIrTe$_4$, the band crossings, associated with WPs, are predominantly governed by the { {Ir-5$d_{z^2}$}} orbital, with a comparatively weaker contribution from the { {Ir-5$d_{xz}$}} orbital, both in the absence and presence of SOC. The corresponding orbital contributions for the SOC-included cases are shown in Fig.\ref{fignew1}. The above analysis refers to the fact that the type-I WP W$_3$ in TaRhTe$_4$, arising from the combined contributions of the { {Rh-4$d_{xz}$ and -4$d_{z^2}$} orbitals of Rh, transforms into a type-II WP 
after replacing  Rh with Ir in TaIrTe$_4$  where Ir-5$d_{z^2}$ orbital predominantly contributes  as shown in Fig.\ref{fig2}(a).  The replacement of Rh by Ir leads to a redistribution of orbital occupancy, accompanying  a strong contribution from  $d_{z^2}$- as compared to $d_{xz}$-bands, near the Fermi level. This orbital reorganization increases the tilting of the WPs, driving a topological phase transition from a hybrid WSM to a type-II WSM under Rh $\to$ Ir without altering the underlying crystal symmetry.  Therefore,  in presence of SOC,  TaXTe$_4$ undergoes a orbital driven topological phase transition from hybrid (TaRhTe$_4$) to type-II (TaIrTe$_4$) WSM when Ir replaces Rh.  This phase transition is primarily governed by the relative contributions of $d_{xz}$ and $d_{z^2}$ orbitals associated with Rh/Ir atoms.}}


\section{Effect of topological phase transition in chirality mediated planar Hall transport}
 \label{sec3}

To explore the effect of topological phase transition from type-II (TaIrTe$_4$) to hybrid (TaRhTe$_4$) WMS in ternary tellurides family,  we investigated planar Hall response which appears as a smoking gun evidence for the chiral anomaly.  The geometrical set up for calculating planar Hall response in TaXTe$_4$ is shown in Fig.\ref{fig4}(a).   In TaXTe$_4$ (X=Rh, Ir),  Ta and X atoms are nestled between two layers of Te atoms which creating a Te–TaX–Te sandwich-like configuration.   As the  van der Waals stacking of layer is along the $z$-axis, therefore we chose the electric and magnetic fields in the $xy$-plane to calculate the planar Hall conductivity. Following the semi-classical Boltzmann transport equation and relaxation time approximation, the planar Hall conductivity $\sigma_{\alpha\beta}$ is found to be  
\cite{PhysRevLett.119.176804, Nandy_2018, Tanay_2018,PhysRevB.93.035116}
\begin{eqnarray}
\sigma_{\alpha \beta}^{\gamma}&\simeq e^{2}& \displaystyle \int\frac{d^{3}k}{(2\pi)^{3}}D\tau\left(-\frac{\partial f_{0}}{\partial \epsilon}\right) 
\Big[ \big(v_{\alpha}+\frac{eB\sin \phi}{\hbar}(\mathbf{\Omega_{k}}\cdot\mathbf{v_{k}})\big) \nonumber \\
&\times &\big(v_{\beta}+\frac{eB\cos \phi}{\hbar}(\mathbf{\Omega_{k}}\cdot\mathbf{v_{k}})\big)\Big]
\label{eq_ehc}
\end{eqnarray}
where $D\equiv D(\mathbf{B,\Omega_{k}})=(1+\frac{e}{\hbar}(\mathbf{B}\cdot \mathbf{\Omega_{k}}))^{-1}$ is the phase space
factor~\cite{Duval_2006} and $\gamma \ne \alpha, \beta$. The Berry curvature and velocity are denoted by $\mathbf{\Omega_{k}}=(\Omega_x,\Omega_y,\Omega_z)$ and  $\mathbf{v_{k}}=(v_x,v_y,v_z)$, respectively. The magnetic field is given by ${\mathbf B}= B \cos \phi \hat{i} + B \sin \phi \hat{j} $, electric field ${\mathbf E}= E \hat{i}$.  The computational details for calculating planar Hall response for both ternary tellurides are described in supplementary materials \cite{SM}.

Figure \ref{fig4}(b) represents the planar Hall conductivity (${\sigma}^z_{xy}$) for both TaIrTe$_4$ and TaRhTe$_4$ with the chemical potential at Fermi level (E$_f$=0 eV).  The planar Hall conductivity enhances due to orbital driven topological phase transition from type-II phase of  TaIrTe$_4$ to hybrid WSM phase of TaRhTe$_4$.  To explore further about the dependency of Hall response on the type and nature of WP in ternary tellurides,  we study planar Hall conductivity at each WP energy for both TaRhTe$_4$ and TaIrTe$_4$ as depicted in Fig.\ref{fig4}(c) and (d), respectively.  In TaRhTe$_4$,  W$_1$ and W$_2$ are of type-II in nature,  whereas W$_3$ is of type-I.  The planar Hall conductivity increases for type-II  WPs compared to type-I WPs in TaRhTe$_4$.  Planar Hall conductivity enhances almost an order of magnitude when the chemical potential E$_f$ is set at the energy of type-II WP i.e., E$_f$=E$_{W_{1,2}}$ in {{ TaRhTe$_4$. The Planar Hall conductivity increases for TaIrTe$_4$ when E$_f$=E$_{W}$ as compared to $E_f=0$ indicating towards the fact that tilting of WPs causes stronger response. It is evident from Fig.\ref{fig4}(c) and (d) that the $d$-orbital contributes significantly in TaXTe$_4$, X=Rh, Ir when the Fermi level is set at WPs rather than at zero energy. }}

This provides a framework for the experimental observations of enhanced  planar Hall response in ternary tellurides.  Here we also report the enhancement of planar Hall conductivity in TaRhTe$_4$ compared to TaIrTe$_4$ which is due to orbital contribution of band crossing below the Fermi level in TaRhTe$_4$ resulting in a topological phase transition.  This augments the   sensitivity of planar Hall coefficients in
response to band topology thus offering a new tool for probing topological phase transitions in  ternary tellurides family.  Anisotropic nature of the effective mass is caused by the anisotropy in the band dispersion which plays significant role in determining the chiral anomaly-mediated planar Hall responses that we study below.

\begin{figure}[ht]
\centering
\includegraphics[width=0.5\textwidth,angle=0]{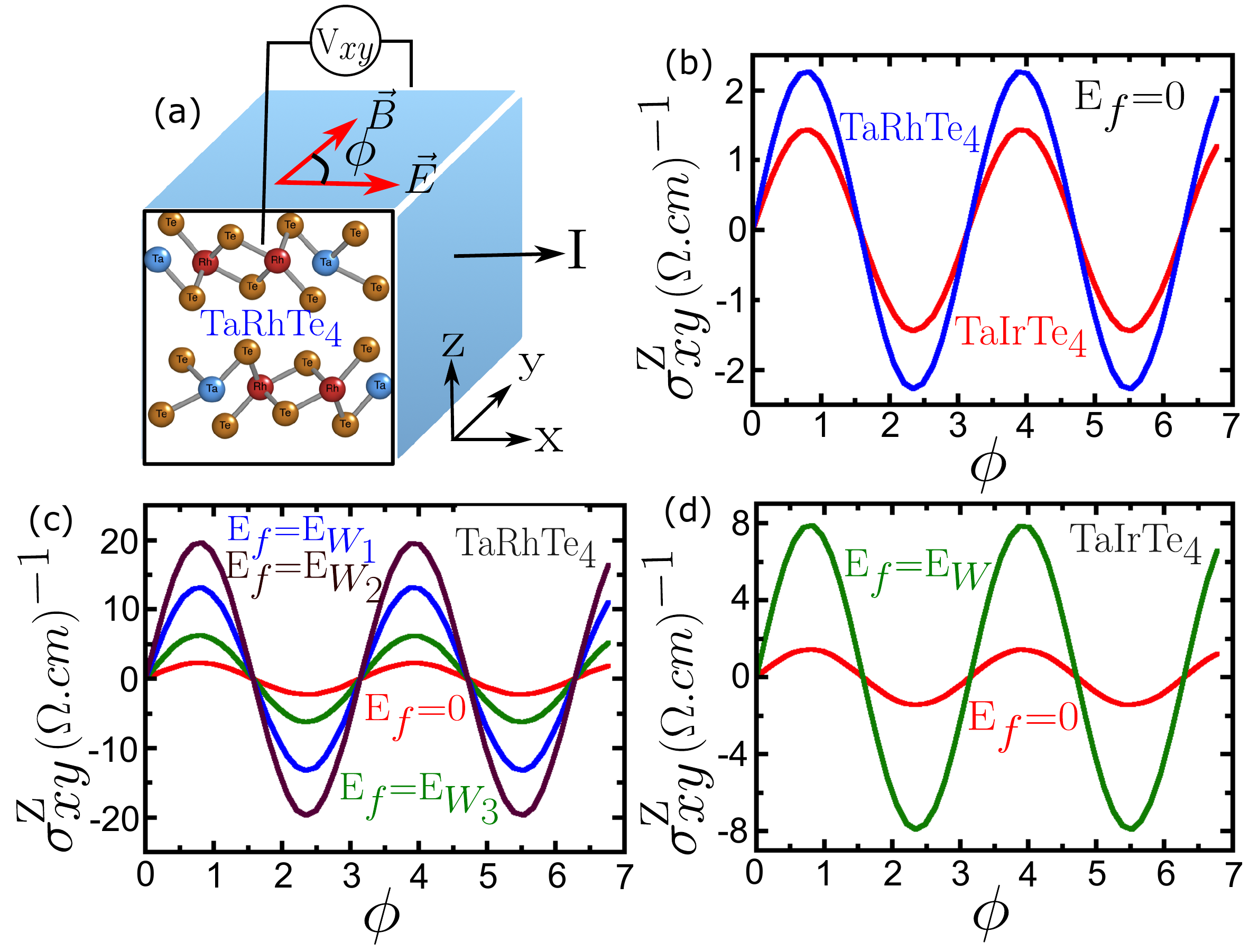}
\caption{(a) Planar Hall setup in TaXTe$_4$ (X=Rh,  Ir) where Ta,  Rh/Ir and Te atoms are shown by blue,  red and brown filled spheres, respectively.   (b) Planar Hall conductivity $\sigma^z_{xy}$, using Eq. (\ref{eq_ehc}), in TaXTe$_4$ with chemical potential at 0 eV.  Planar Hall  conductivity $\sigma^z_{xy}$, using Eq. (\ref{eq_ehc}), for TaRhTe$_4$ and TaIrTe$_4$ are shown in (c)  and (d), respectively with chemical potential kept  at WP energies. }
    \label{fig4}
\end{figure}

\begin{figure*}[ht]
\centering
\includegraphics[width=0.9\textwidth,angle=0]{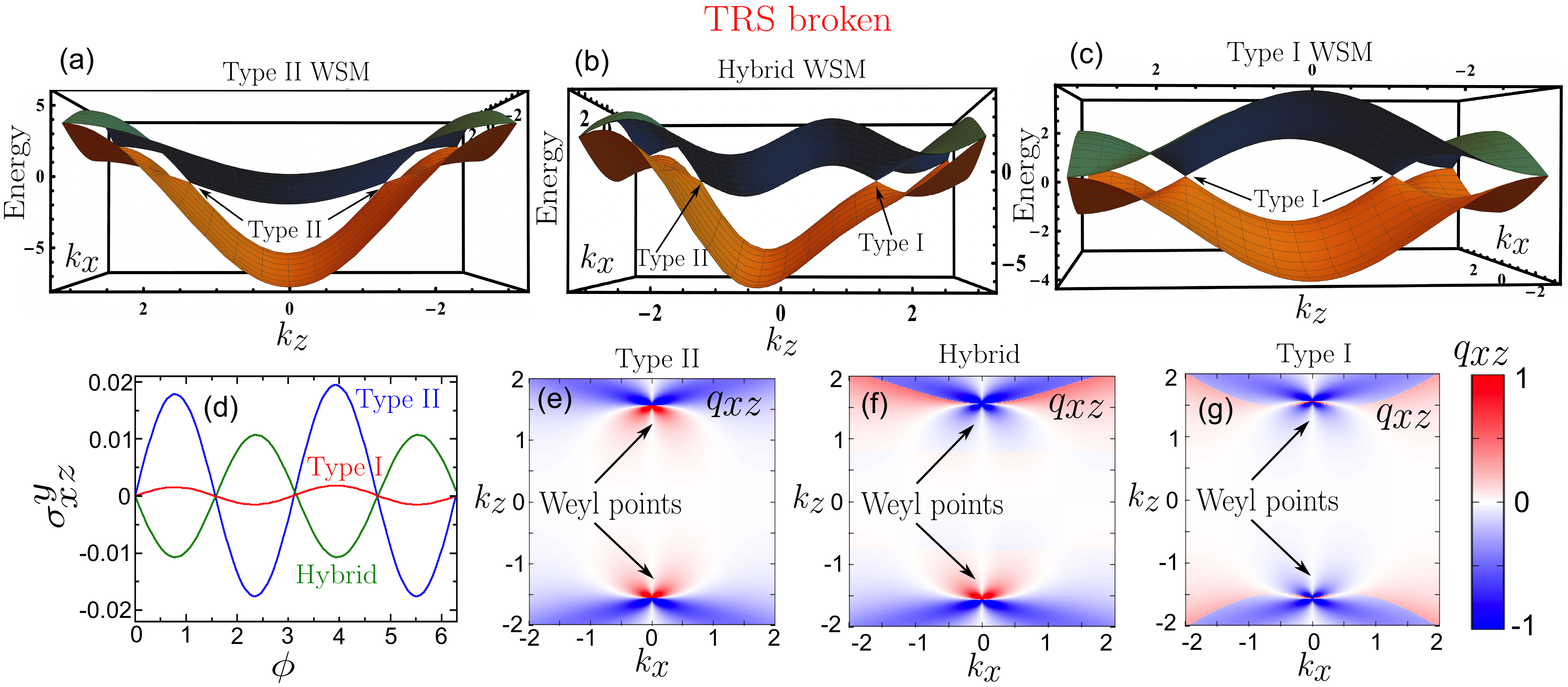}
\caption{Bulk band dispersion of time reversal symmetry broken  WSM tight-binding model given in Eq. (\ref{TRS-broken}), are shown in (a) for type-II phase (with parameters $t_z$=2.0,  $t_1$=1.9,  $t_2$=0, $\phi_1$=$\pi$, $\phi_2$=$\frac{\pi}{2}$) (b) hybrid phase  (with parameters $t_z$=2.0,  $t_1$=1.0,  $t_2$=0.5, $\phi_1$=$\pi$, $\phi_2$=$\frac{\pi}{2}$) and (c) type-I phase (with parameters $t_z$=2.0,  $t_1$=0.1,  $t_2$=0, $\phi_1$=$\pi$, $\phi_2$=$\frac{\pi}{2}$).  (d) Planar Hall transport $\sigma^Z_{xy}$, using Eq. (\ref{eq_ehc}), are shown for type-I,  hybrid and type-II phases. We consider $\mu=0.5$ for all our calculations for better of the results.  Velocity modulated off-diagonal effective masses ($q_{xz}=v_x v_z M_{xz}$) are shown  in (e) type-II, (f) hybrid and (g) type-I phases of WSM model (\ref{TRS-broken})   where $M_{xz}=\frac{1}{{\hbar}^2} \frac{{d^2}E}{dk_x dk_z} \frac{1}{|v_x| |v_z|}$. 
     }
    \label{fig5}
\end{figure*}


\begin{figure}[ht]
\centering
\includegraphics[width=0.5\textwidth,angle=0]{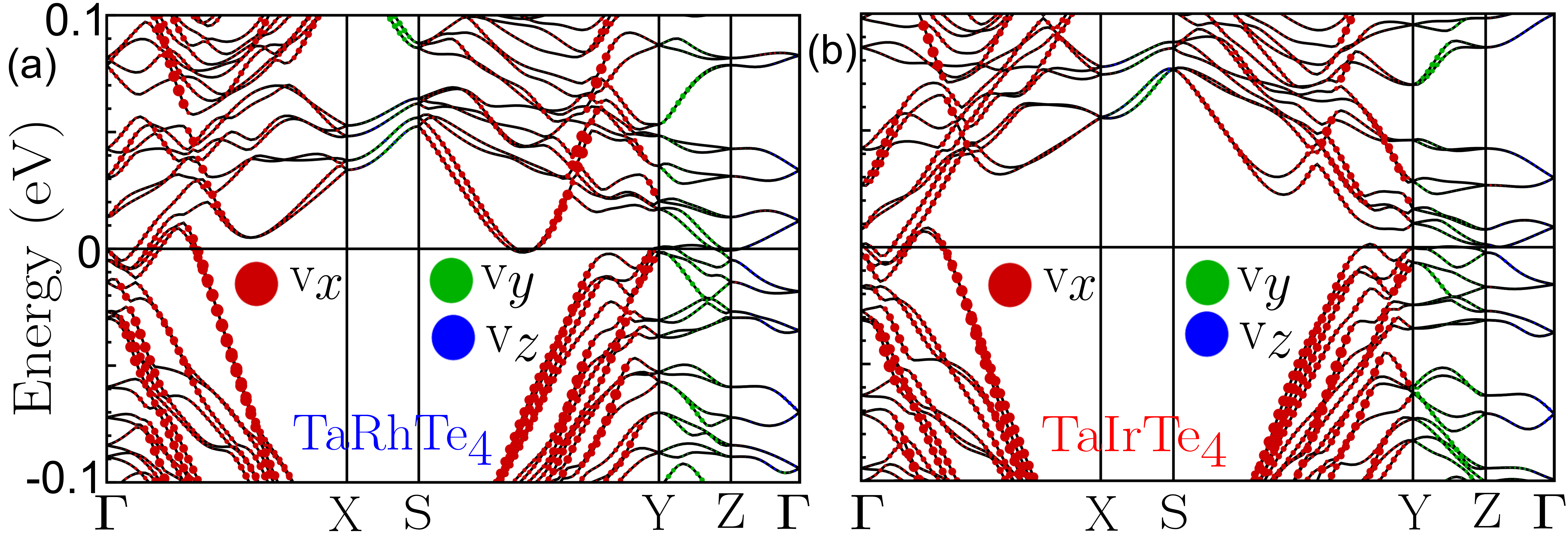}
\caption{Velocity profiles of individual bands for TaRhTe$_4$, and  TaIrTe$_4$ are shown in (a),  and (b), respectively.
}
    \label{fig6}
\end{figure}

\section{Role of velocity modulated off-diagonal effective mass on planar Hall response}
\label{sec4}

{{ To explore the dependency of planar Hall response on the nature of WP,  we construct a tight-binding model Hamiltonian mimicking both the TRS broken and TRS invariant WSM \cite{SM}. At the outset, we note that PHE for TRS invariant and TRS broken WSM may not vary substantially \cite{PhysRevB.107.075131}. We therefore restrict our model calculation here on a TRS broken WSM for a better understanding of material's results in an qualitative manner. The effect of tilt on planar Hall response of a TRS invariant WSM is described in the supplementary materials \cite{SM}. 
The Hamiltonian for TRS broken WSM is given by $H=N_0 \sigma_0 + N_x \sigma_x + N_y \sigma_y+ N_z \sigma_z  $ }}\cite{PhysRevB.94.121105,PhysRevB.105.214307} where
\begin{eqnarray}
 N_0&=&2 t_1 \cos(\phi_1 - k_z) +  2 t_2 \cos(\phi_2 - 2 k_z), \nonumber \\
N_x&=&t \sin k_x,~~ N_y=t \sin k_y, ~~~\rm{and}\nonumber \\
 N_z&=& t_z \cos k_z -m_z +t_0(2-\cos k_x -\cos k_y).
\label{TRS-broken}
 \end{eqnarray}
Here, the $N_0$ term is tailor-made to modulate the the energies as well as tilt of the WPs. Note that $t_{1(2)}$ represents the first (second) nearest-neighbour hopping while $\phi_{1(2)}$ denotes their phases. The phase difference $\phi_1 \ne \phi_2$ is responsible for causing the hybrid nature of the WPs while  the  position and chirality of the WPs are  determined by the $N_{x,y,z}$ terms. The WPs are found to be present at ${\bm k}=(0,0, sk_0)$ with 
\begin{equation}
\cos (s k_0)=\frac{t_0}{t_z}\bigl[ \frac{m_z}{t_0}+\cos k_x +\cos k_y -2\bigr]
\label{eq_lm1}
\end{equation}
and $s=\pm$. Expanding around the WPs, the low-energy  Hamiltonian is given by  $H\approx 2 k_z(t_1 \sin(\phi_1 - s k_0) + 2 t_2 \sin(\phi_2 - 2 s k_0) ) \sigma_0+ t( \sigma_x k_x +  \sigma_y k_y)+ s t_z  \sigma_z k_z \sin k_0$. The energy is given by $E= 2 k_z(t_1 \sin(\phi_1 - s k_0) + 2 t_2 \sin(\phi_2 - 2 s k_0) ) \pm \sqrt{t^2k_x^2 +t^2k_y^2+ t_z^2 \sin^2 k_0k_z^2}$. A careful analysis suggests that the  WP at $k_z=-\pi/2$ is of type-I (type-II) while the right WP at $k_z=+\pi/2$ belongs to type-II (type-I) with $k_0=\pi/2$. The important point is that the above model allows one to obtain type-I, type-II and
hybrid phases by appropriately tuning the parameters $t_{1,2}$ and
$\phi_{1,2}$. For example, given $(\phi_1,\phi_2)=(\pi,\pi/2)$, both the WPs becomes type-I (type-II) for $t_2<-0.5 t_1 +0.5$ ($t_2<0.5 t_1 -0.5$ and $t_2>0.5 t_1 +0.5$) while the hybrid phase appears when  $t_2 < 0.5t_1 + 0.5$,
$t_2 > -0.5 + 0.5t_1$ and $t_2 > -0.5 t_1 + 0.5$.

We examine the planar Hall conductivity Eq. (\ref{eq_ehc}) for the tight-binding model of TRS broken WSM given in Eq. (\ref{TRS-broken}) to study the effect of tilt.We first show the dispersion associated with the type-II, hybrid and type-I phases in Figs. \ref{fig5} (a,b,c), respectively while the tilt is present along $k_z$ direction. We now examine the planar Hall conductivity { {$\sigma_{xz}^y$}} with the relative angle between the electric and magnetic field. Under a suitable parameter window, we find the response of type-II WSMs is the most significant among all three phases while the magnitude of the response associated with type-I WSMs is the lowest, see Fig. \ref{fig5} (d) for variations in $\sigma_{xz}^y$  for TRS broken type-I, hybrid and type-II model Hamiltonian. This is what is also observed in our material's simulation of TaRhTe$_4$, as shown in Fig. \ref{fig4}(c).  Therefore, a tight-binding model with tilt along $k_z$ is able to mimic the finding of a material where the tilt is along $k_x$.

Interestingly, we find a nice correlation between the planar Hall conductivity and  profile of $q_{xz}=v_x v_z M_{xz}$  in the $k_x$-$k_z$ plane containing the effect of the tilt along $k_z$ direction in the model Hamiltonian   \cite{PhysRevB.105.205207,  yu2005fundamentals}.  {{Here $M_{xz}$ represents the off-diagonal effective mass and $q_{xz}$ is the velocity modulated off-diagonal effective mass where $M_{xz}=\frac{1}{{\hbar}^2} \frac{{d^2}E}{dk_x dk_z} \frac{1}{|v_x| |v_z|}$}}. This is shown in Fig. \ref{fig5} (e,f,g) for type-II, hybrid and type-I, respectively, where {{$q_{xz}$}} exhibits markedly different behavior around the WPs.  To be more precise, the quantity $q_{xz}$ shows identical profile with negative (positive) signs around both the WPs in  type-I (type-II) phase while it acquires an opposite sign profile between the  WPs for hybrid phase. This behavior can be  related to the maximum  (minimum) magnitude of planar Hall conductivity in the type-II (type-I) phase while an intermediate magnitude is observed for the hybrid case where {{$q_{xz}$}}  changes sign between the WPs.

{{In other words, the profile of $q_{xz}$ around two WPs look qualitatively different in hybrid phase while the profiles are qualitatively identical across the WPs in type-I and II phases. The parity is determined by relative change in the $q_{xz}$ profile around two WPs. There is always a sign change of $q_{xz}$ around both the type-II WPs while there is no sign reversal of $q_{xz}$ around type-I WPs. This leads to the identical parity of WPs as far as the type-I and type-II phases are concerned. This is markedly different when it comes to hybrid phase in which the sign reversal of $q_{xz}$ happens (does not happen) around type-II (type-I) WP. This corresponds to the fact that parity of WPs are different in hybrid phases.    The phase of the response $\sigma^y_{xz}$ changes in hybrid phase as compared to the type-I and II phases. This can naively explain the response $\sigma^y_{xz}$ as shown in Fig. \ref{fig5} (d).}} Note that the behavior of the model  only provides a naive indication of the planar Hall conductivity profile that is observed in the material.

 In the below we provide a tentative connection between the Berry phase and off-diagonal effective mass to strengthen our argument.  As a whole, the contributions coming from Berry curvature and velocity
can be naively mimicked by the off-diagonal terms of the effective mass \cite{PhysRevB.105.205207}. 
There exists an indirect connection between Berry curvature and effective mass as discussed below.  
For the multi-band case, the effective mass acquires an inter-band correction term in addition to the intra-band conventional term. The effective-mass correction can be viewed as a gap-weighted sum of those symmetric quantum geometric tensor contributions \cite{yu2005fundamentals} while the anti-symmetric part of the quantum geometric tensor is Berry curvature.  
In the case of the material $\sigma^z_{xy}$, one can naively anticipate that $(\mathbf{ \Omega_k}\cdot \mathbf{v_k })^2$ in Eq. (\ref{eq_ehc}) is connected with {{ $q_{xz}=v_x v_z M_{xz}= {\rm sgn}(v_x) {\rm sgn}(v_z) (\partial v_x/\partial k_z)
$}} as the tilt is  primarily present along $k_x$ direction. {{Note that the  $\sigma^a_{bc}$ is related to $v_b v_c M_{bc}$ and there exist a correlation between the  velocity modulated off-diagonal effective masses and the corresponding Hall coefficients.}}

The above analysis refers to the fact that  the velocity term $v_x$ acquiring a dominant contribution when the quantity $q_{xb}$, with $b=y,z$, is studied for bands around the WPs.  In order to mimic the effect of the tilt in the materials TaXTe$_4$, we show the  velocity profiles of each of the bands where $v_x$ dominates over $y$ and $z$ contributions of velocities because of strong tilting of WPs along $k_x$ direction, see Fig. \ref{fig6} (a,b). This gives a hint that the relative magnitude of  $\sigma^z_{xy}$ in  TaXTe$_4$ depends on the tilt, possibly through a $q_{xy}$-like term. Given a correlation between the behavior of $q_{xz}=v_x v_z M_{xz}$ around the WPs and $\sigma^y_{xz}$ in the model, we can comment that $v_x$ component in $q_{xy}=v_x v_y M_{xy}$ plays substantial role to tune the behavior of $\sigma^z_{xy}$ in  TaXTe$_4$.  Hence our findings on the tight-binding model provide a naive understanding of the numerical results obtained for the material.

\begin{figure}[ht]
\centering
\includegraphics[width=0.3\textwidth,angle=0]{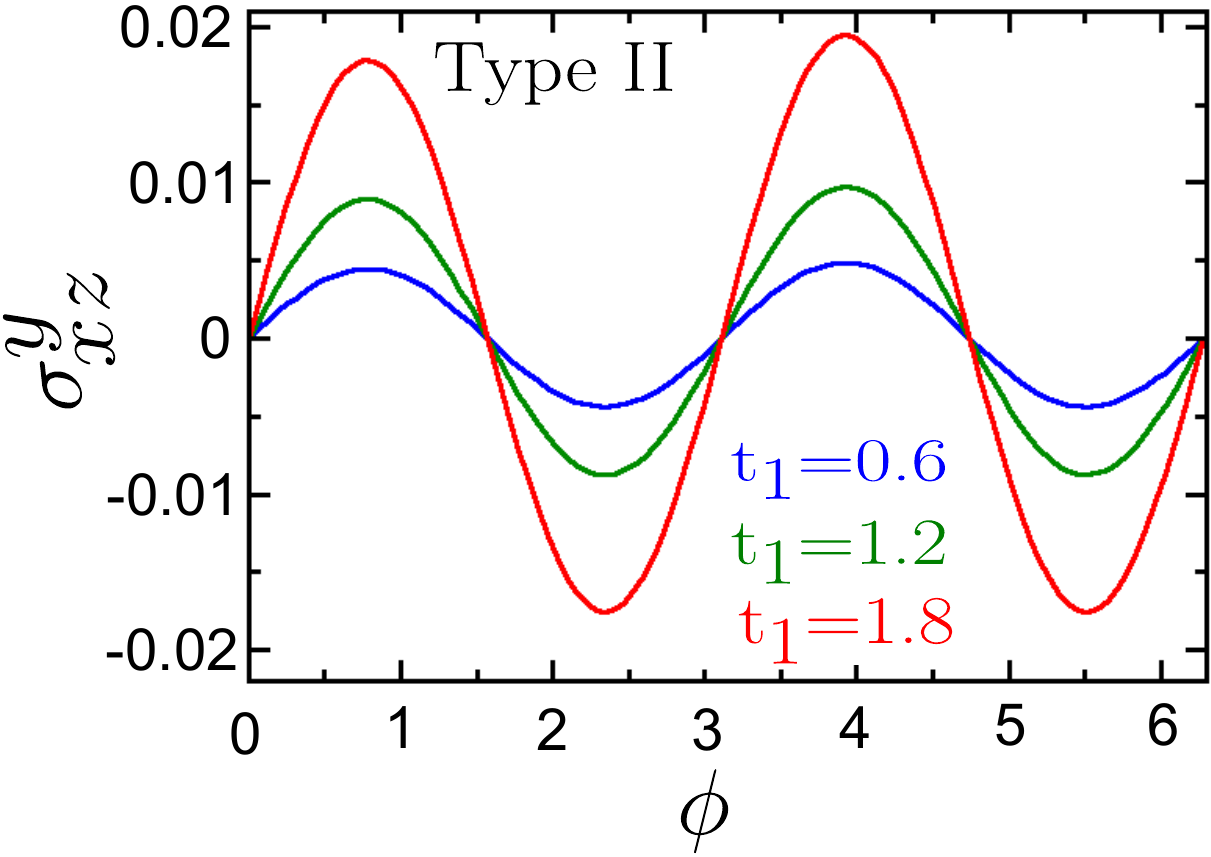}
\caption{Planar Hall transport in TRS broken type II WSM with tilting parameter t$_1$.}
    \label{fignew3}
\end{figure}

{ {In a type-I WSM, the planar Hall conductivity $\sigma_{xy}^z$ receives contributions solely from electron pockets.  In contrast, for a type-II WSM,  both electron and hole pockets contribute to $\sigma_{xy}^z$. Notably, the contribution from hole pockets increases with the tilt parameter of the WP under consideration, leading to an enhancement of the planar Hall conductivity. It can be anticipated that planar Hall conductivity $\sigma_{xz}^y$  systematically increases as the tilt parameter in the time-reversal-symmetry-broken type-II WSM model increases as shown in Fig. \ref{fignew3}.  In general, this can effectively explain the  enhancement of the planar Hall conductivity in hybrid phases as compared to other phases in TaXTe$_4$ materials.}}

\section{Conclusion}
 \label{sec5}

We report the effect of SOC and orbital-driven topological phase transition in ternary tellurides TaXTe$_4$ (X=Rh, Ir).  We show that TaRhTe$_4$ hosts WPs of both type-I and type-II,  whereas TaIrTe$_4$ has only type-II (type-I) WPs with SOC (without SOC).  All WPs in TaXTe$_4$  lie in the $k_z=0$ plane but remain well-separated in energy and momentum space.  TaIrTe$_4$ appears as type-I WSM and under the application of SOC  it converts from a  type-I WSM to a type-II WSM. {{ With SOC,  in hybrid WSM
TaRhTe$_4$ [type-II WSM TaIrTe$_4$],
d$_{xz}$ and d$_{z^2}$ [d$_{z^2}$] orbitals  contribute significantly to the WPs. In TaXTe$_4$, 
replacing X=Rh with isoelectronic and isostructural X=Ir atom,   orbital-driven topological phase transition occurs from a hybrid to type-II where there is a change in the relative contribution from  
d$_{xz}$ and d$_{z^2}$ orbitals to the WPs. }}

The PHE in WSM has been proposed as a key feature of chiral anomaly and we report enhancement  of PHE due to orbital-driven topological phase  transition in TaXTe$_4$ from ab initio calculations.  The resulting Fermi arcs connecting WPs of opposite chirality have also been identified, which have interesting consequences for transport. {{ The magnitude of planar Hall conductivity differs noticeably between TaRhTe$_4$ and TaIrTe$_4$.  We further investigate the role of velocity modulated off-diagonal effective mass on the planar Hall response from tight-binding model calculations to understand the materials behavior.  Our study explores that orbital contribution has large effect on topological phases in ternary tellurides than SOC.  We believe,  therefore,  our work triggers a fertile background for further exploration of topological transport phenomena in ternary tellurides.}}


\section{Acknowledgement}

B.S. acknowledges discussions with Jeroen van den Brink at an early stage of this work. This work was supported by the Prime Minister’s Early Career Research Grant (PMECRG) of the Anusandhan National Research Foundation (Grant No. ANRF/ECRG/2024/005021/PMS). The author thanks the IFW Dresden cluster and Ulrike Nitzsche for technical assistance.



%

\end{document}